# Well-Definedness and Semantic Type-Checking in the Nested Relational Calculus and XQuery

## Extended Abstract

Jan Van den Bussche[1], Dirk Van Gucht[2*], and Stijn Vansummeren[1**]

[1] Limburgs Universitair Centrum
Diepenbeek, Belgium
jan.vandenbussche@luc.ac.be, stijn.vansummeren@luc.ac.be
[2] Indiana University
Bloomington, Indiana, USA
vgucht@cs.indiana.edu

**Abstract.** Two natural decision problems regarding the XML query language XQuery are well-definedness and semantic type-checking. We study these problems in the setting of a relational fragment of XQuery. We show that well-definedness and semantic type-checking are undecidable, even in the positive-existential case. Nevertheless, for a "pure" variant of XQuery, in which no identification is made between an item and the singleton containing that item, the problems become decidable. We also consider the analogous problems in the setting of the nested relational calculus.

* The author is supported by NSF Grant IIS-0082407
** Research Assistant of the Fund for Scientific Research - Flanders (Belgium)

## 1 Introduction

Much attention has been paid recently to XQuery, the XML query language currently under development by the World Wide Web Consortium [5, 9]. Unlike in traditional query languages, expressions in XQuery can have an undefined meaning (i.e., these expressions produce a run-time error). As an example, consider the following variation on one of the XQuery use cases [7]:

```
<bib> {
  for $b in $bib/book
  where $b/publisher = "Springer-Verlag"
  return element{$b/author}{$b/title}
 } </bib>
```

This expression should create for each book published by Springer-Verlag a node whose name equals the author of the book, and whose child is the title of the book. If there is a book with more than one `author` node however, then the result of this expression is undefined because the first argument to the element constructor must be a singleton list.

This leads us to the natural question whether we can solve the *well-definedness problem* for XQuery: given an expression and an input type, check whether the semantics of the expression is defined for all inputs adhering to the input type. This problem is undecidable for any computationally complete programming language, and hence also for XQuery. Following good programming language practice, XQuery therefore is equipped with a static type system (based on XML Schema [4, 18]) which ensures "type safety" in the sense that every expression which passes the type system's tests is guaranteed to be well-defined. Due to the undecidability of the well-definedness problem, such type systems are necessarily incomplete, i.e., there are expressions which are well-defined, but not well-typed.

Can we find fragments of XQuery for which the well-definedness problem is decidable? In this paper we will study *Relational XQuery* (RX), a set-based fragment of XQuery where we omit recursive functions, only allow the child axis, take a value-based point of view (i.e., we ignore node identity), and use a type system similar to that of the nested relational or complex object data model [1, 6, 19]. We regard RX as the "first-order database fragment" of XQuery.

Even for RX, the well-definedness problem is still undecidable, due to two features which allow us to simulate the relational algebra: quantified expressions and type switches. Surprisingly, however, well-definedness remains undecidable for RX without these features, which we call positive-existential RX or PERX for short.

The core difficulty here is due to the fact that in the XQuery data model an item is identified with the singleton containing that item [11]. In a set-based model this identification becomes difficult to analyze, since $\{i, j\}$ is a singleton if and only if $i = j$. Since, as shown in the example above, there are expressions which are undefined on non-singleton inputs, this implies that in order to solve the well-definedness problem, one also needs to solve the equivalence problem. Indeed, we will see that the equivalence problem for PERX is undecidable.

Nevertheless, for a "pure" variant of PERX, in which no identification is made between an item and the singleton containing that item, well-definedness becomes decidable. We actually prove this result not for pure PERX itself, but for PENRC: the positive-existential fragment of the *nested relational calculus* [6, 19], which is well-known from the complex object data model, and whose well-definedness problem is interesting in its own right.

All our results hold not only for well-definedness, but also for *semantic type-checking*: given an expression, an input type and an output type, check whether the expression always returns outputs adhering to the output type on inputs adhering to the input type.

In the main body of the paper we will work in a set-based data model. Considering that the real XML data model is list-based, at the end of the paper we will discuss how and if our results transfer to a list-based or bag-based setting.

*Related work* The semantic type-checking problem has already been studied extensively in XML-related query languages [2, 3, 13–15, 17]. In particular, our setting closely resembles that of Alon et al. [2, 3] who, like us, study the problem in the presence of data values. In particular they have shown that (un)decidability depends on the expressiveness of both the query language and the type system. While the query language of Alon et al. can simulate PERX, our results do not follow immediately from theirs, since their type system is incompatible with ours [16].

## 2 Relational XQuery

In what follows we will need to define various query languages. In some definitions it will help to talk abstractly about a query language. To this end, we define a *query language* $Q$ as a tuple $(V, T, E, [\![.]\!])$ where $V$ is a set of *values*; $T$ is a set of *types*; $E$ is a set of *expressions*; and $[\![.]\!]$ is the interpretation function giving a semantics to types and expressions. The set $V$ is also referred to as the data model.

We assume to be given an infinite set $\mathcal{X} = \{x, y, \dots\}$ of *variables*. Every expression $e$ has associated with it a finite set $FV(e) \subseteq \mathcal{X}$ of *free variables*. An *environment* on $e$ is a function $\sigma : FV(e) \to V$ which associates to each $x \in FV(e)$ a value $\sigma(x) \in V$. A *type assignment* on $e$ is a function $\Gamma : FV(e) \to T$ which associates to each $x \in FV(e)$ a type $\Gamma(x) \in T$. If $\rho$ is an environment (or a type assignment), and $v$ is a value (respectively a type), then we write $x : v, \rho$ for the environment (respectively type assignment) $\rho'$ with domain $dom(\rho) \cup \{x\}$ such that $\rho'(x) = v$ and $\rho'(y) = \rho(y)$ for $y \neq x$. Intuitively, environments describe the input to expressions, and type assignments describe their type.

Every type $\tau$ is associated with a set $[\![\tau]\!]$ of values. An environment $\sigma$ is *compatible with* a type assignment $\Gamma$, denoted by $\sigma \in \Gamma$, if they have the same domain and $\sigma(x) \in [\![\Gamma(x)]\!]$ for all $x$. Every expression $e$ has associated with it a (possibly partial) computable function $[\![e]\!]$ which associates environments on $FV(e)$ to values in $V$. We call $[\![e]\!]$ the *semantics* of $e$.

In order not to burden our notation we will identify types and expressions with their respective interpretations, and write for example $e(\sigma)$ for $[\![e]\!](\sigma)$.

### 2.1 Relational XQuery data model

In this section we define a set-based fragment of the XQuery data model [11] called the *Relational XQuery (RX) data model.* We take a value-based point of view (i.e., we ignore node identity), focus on data values, element nodes and data nodes (known as text nodes in XQuery), and abstract away from the other features in the XQuery data model such as attributes.

We assume to be given a recursively enumerable set $\mathcal{A} = \{a, b, \dots\}$ of *atoms.* An *item* is an atom or a *node.* A node is either an *element node* $\langle a : N \rangle$ or a *data node* $\langle a \rangle$, where $a \in \mathcal{A}$ and $N$ is a finite set of nodes ($N$ is called the content of the element node). An *RX-value*, finally, is any finite set of items.

An *RX-type* $\tau$ is a term generated by the following grammar:

$$\begin{aligned} \tau &::= \mathbf{coll}(\iota) \mid \mathbf{single}(\iota) \\ \iota &::= \mathbf{atom} \mid \nu \mid \iota \cup \iota \\ \nu &::= \mathbf{data} \mid \mathbf{elem}(\gamma) \mid \nu \cup \nu \\ \gamma &::= \mathbf{coll}(\nu) \mid \mathbf{single}(\nu) \end{aligned}$$

Here, $\tau$ ranges over types, $\iota$ ranges over item types, $\nu$ ranges over node types, and $\gamma$ ranges over node content types. An RX-type denotes a set of RX-values:

- **data** denotes the set of all data nodes,
- **elem**($\gamma$) denotes the set of all element nodes $\langle a : N \rangle$ for which $N$ is a finite set over the denotation of $\gamma$,
- **atom** denotes the set $\mathcal{A}$ of all atoms,
- $\iota_1 \cup \iota_2$ denotes the union of the denotations of $\iota_1$ and $\iota_2$,
- **coll**($\iota$) denotes the set of all finite sets over the denotation of $\iota$, and
- **single**($\iota$) denotes the set of all singletons over the denotation of $\iota$.

Note that every $\gamma$ is also a $\tau$, and hence the denotation of terms produced by $\gamma$ is subsumed in the definition above.

An *RX-kind* $\kappa$ is a term generated by the following grammar:

$$\kappa ::= \mathbf{atom} \mid \mathbf{data} \mid \mathbf{elem} \mid \kappa \cup \kappa$$

An RX-kind denotes a set of items, which can be the set of all atoms, the set of all data nodes, the set of all element nodes, or the union of the denotations of two kinds.

*Discussion* The type system we have defined above is quite simple. Types merely indicate the many-or-one cardinality of a value, and the kinds of items that can appear in it. Only values of a fixed maximal nesting height can be described in our type system. This is justified because the expressions in the XQuery fragment RX we will work with in this paper can look only a fixed number of nesting levels down anyway. Also, it is a public secret that most XML documents in practice have nesting heights at most five or six, and that unbounded-depth nesting is not needed for many XML data processing tasks.

The presence of the **single** type constructor is justified by the fact that an item $i$ is identified with the singleton set $\{i\}$ in the XQuery data model [11]. Consequently, an XQuery expression in which the input is always expected to be a string actually receives singleton strings as inputs. Its input type would therefore be **single**(**atom**) in our setting.

Our types also do not specify anything about the names of element nodes, but this is an omission for the sake of simplicity; we could have added node types of the form $\mathbf{elem}_a(\gamma)$, with $a$ the atom that must be the name of the element node, without sacrificing any of the results we present in this paper.

### 2.2 Relational XQuery syntax and semantics

A *Relational XQuery expression* is an expression generated by the following grammar:

$$\begin{aligned} e ::= {} & x \\ \mid {} & \mathit{text}\{e\} \mid \mathit{elem}\{e\}\{e\} \mid \mathit{data}(e) \mid \mathit{name}(e) \mid \mathit{children}(e) \\ \mid {} & () \mid e, e \mid \mathit{for}\ x : \kappa\ \mathit{in}\ e\ \mathit{return}\ e \\ \mid {} & \mathit{if}\ e\ \mathit{eq}\ e\ \mathit{then}\ e\ \mathit{else}\ e \mid \mathit{if}\ e = \emptyset\ \mathit{then}\ e\ \mathit{else}\ e \mid \mathit{if}\ e \in \tau\ \mathit{then}\ e\ \mathit{else}\ e \end{aligned}$$

Here, $e$ ranges over RX-expressions, $x$ ranges over variables, $\tau$ ranges over RX-types and $\kappa$ ranges over RX-kinds. The *free variables* of $e$ are defined in the usual way, and will be denoted by $FV(e)$.

The semantics of RX is parameterized by two "oracle" functions:

- *content*, which maps element nodes to atoms; and
- *concat*, which maps finite sets of atoms to atoms.

We further define the following (partial) functions on values:

- $\mathit{data}(v) = \{a \mid a \in v\} \cup \{a \mid \langle a \rangle \in v\} \cup \{\mathit{content}(\langle a : N \rangle) \mid \langle a : N \rangle \in v\}$,
- $\mathit{name}(v)$, which is $\{a\}$ if $v$ is a singleton element node $\{\langle a : N \rangle\}$; $\mathit{concat}(v)$ if $v$ is empty; and undefined otherwise.
- $\mathit{children}(v)$, which is undefined if there is some atom in $v$, and otherwise returns
$$\bigcup\{N \mid \langle a : N \rangle \in v\}.$$
- $\mathit{construct}(v, w)$ which is undefined if $\mathit{data}(v)$ is not a singleton atom $\{a\}$; and returns $\langle a : N \rangle$ otherwise, where $N$ is obtained from $w$ by replacing every atom in $w$ by a corresponding data node:
$$N = \{\langle a \rangle \mid a \in w\} \cup \{i \mid i \in w,\ i \text{ is a node}\}.$$

Let $e$ be an RX-expression and let $\sigma$ be an RX-environment on $e$.[3] The *semantics* $e(\sigma)$ of $e$ under $\sigma$ can now be inductively defined as follows:

$$
\begin{aligned}
x(\sigma) &= \sigma(x)\\
\mathit{text}\{e\} &= \{\langle \mathit{concat}(\mathit{data}(e(\sigma)))\rangle\}\\
\mathit{elem}\{e_1\}\{e_2\}(\sigma) &= \{\mathit{construct}(e_1(\sigma), e_2(\sigma))\}\\
\mathit{data}(e)(\sigma) &= \mathit{data}(e(\sigma))\\
\mathit{name}(e)(\sigma) &= \mathit{name}(e(\sigma))\\
\mathit{children}(e)(\sigma) &= \mathit{children}(e(\sigma))\\
()(\sigma) &= \emptyset\\
e_1, e_2(\sigma) &= e_1(\sigma) \cup e_2(\sigma)\\
\mathit{for}\ x : \kappa\ \mathit{in}\ e_1\ \mathit{return}\ e_2 &= \bigcup\{e_2(x : \{i\}, \sigma) \mid i \in e_1(\sigma) \cap \kappa\}\\
(\mathit{if}\ e_1\ \mathit{eq}\ e_2\ \mathit{then}\ e_3\ \mathit{else}\ e_4)(\sigma) &= \begin{cases} e_3(\sigma) & \text{if } \mathit{data}(e_1(\sigma)) = \mathit{data}(e_2(\sigma)) = \{a\},\\ & \text{with } a \text{ an atom}\\ e_4(\sigma) & \text{if } \mathit{data}(e_1(\sigma)) = \{a\},\ \mathit{data}(e_2(\sigma)) = \{b\},\\ & \text{with } a, b \text{ atoms, } a \neq b \end{cases}\\
(\mathit{if}\ e_1 = \emptyset\ \mathit{then}\ e_2\ \mathit{else}\ e_3)(\sigma) &= \begin{cases} e_2(\sigma) & \text{if } e_1(\sigma) = \emptyset\\ e_3(\sigma) & \text{otherwise} \end{cases}\\
(\mathit{if}\ e_1 \in \tau\ \mathit{then}\ e_2\ \mathit{else}\ e_3)(\sigma) &= \begin{cases} e_2(\sigma) & \text{if } e_1(\sigma) \in \tau\\ e_3(\sigma) & \text{otherwise} \end{cases}
\end{aligned}
$$

Note that $e(\sigma)$ is not necessarily defined: this models the situations in which XQuery expression evaluation produces a run-time error. Specifically, $e(\sigma)$ can become undefined for the following reasons:

- $e = \mathit{elem}\{e_1\}\{e_2\}$, and $\mathit{data}(e_1(\sigma))$ is not a singleton atom. (This can only happen if $e_1(\sigma)$ is not a singleton.)
- $e = \mathit{name}(e')$, and $e'(\sigma)$ is not a singleton element node.
- $e = \mathit{children}(e')$, and $e'(\sigma)$ contains an atom.
- $e = \mathit{if}\ e_1\ \mathit{eq}\ e_2\ \mathit{then}\ e_3\ \mathit{else}\ e_4$, and $\mathit{data}(e_1(\sigma))$ is not a singleton atom, or $\mathit{data}(e_2(\sigma))$ is not a singleton atom. (This can only happen if $e_1(\sigma)$ respectively $e_2(\sigma)$ is not a singleton.)

*Relation to XQuery* The RX query language corresponds to a set-based version of XQuery [5, 9] where we have omitted recursive functions, literals, arithmetic expressions, generalized and order comparisons, and only allow the children axis. We have replaced XQuery quantified expressions by the emptiness test (which is equivalent in expressive power), and have moved kind tests from XQuery step expressions to the "for" expression. As an example, the XQuery step expression $\$x/\mathit{child} :: \mathit{text}()$ can be expressed in RX as

$$\mathit{for}\ z : \mathbf{data}\ \ \mathit{in}\ \ \mathit{children}(x)\ \ \mathit{return}\ z.$$

[3] Recall from the beginning of this section that $\sigma$ assigns an RX-value to each free variable of $e$.

The "oracle" functions *concat* and *content* model features which are present in XQuery, but which are clumsy to take into account in our data model. For example *name* applied to the empty set returns the empty string in XQuery. Furthermore, applying *data* to a singleton element node in XQuery returns the "string content" of the node. This is (roughly speaking) a concatenation of all atoms (converted to strings) encountered in a depth-first left-to-right traversal of the node's content.

## 3 Well-definedness and semantic type-checking

As we have noted in Section 2.2, the semantics $e(\sigma)$ of RX-expression $e$ under environment $\sigma$ can be undefined. This leads us to the following definition.

**Definition 1.** *The well-definedness problem for a query language $Q$ consist of checking, given a $Q$-expression $e$ and a $Q$-type assignment $\Gamma$ on $e$: whether $e(\sigma)$ is defined for every $\sigma \in \Gamma$. In this case we say that $e$ is well-defined under $\Gamma$.*

A problem which is related to well-definedness is the *semantic type-checking* problem:

**Definition 2.** *The* semantic type-checking *problem for a query language $Q$ consist of checking, given a $Q$-expression $e$, a $Q$-type assignment $\Gamma$ on $e$ such that $e$ is well-defined under $\Gamma$, and a $Q$-type $\tau$: whether $e(\sigma) \in \tau$ for every $\sigma \in \Gamma$. In this case we say that $\tau$ is an output type for $e$ under $\Gamma$.*

## 4 Undecidability results

We will show that well-definedness for RX is undecidable, even for a quite restricted fragment. Our results do not depend on the particular interpretation given to the oracle functions *concat* and *content*.

Let us begin by defining $\mathrm{RX}^-$ as the fragment of RX where

- we disallow data node construction expressions of the form $\mathit{text}\{e\}$;
- we disallow data extraction expressions of the form $\mathit{data}(e)$; and
- we disallow kind tests, or equivalently, we only allow the use of the single "universal" kind $\mathbf{atom} \cup \mathbf{data} \cup \mathbf{elem}$.

An $\mathrm{RX}^-$-expression $e$ is *positive existential* if it does not contain *emptiness tests* of the form $\mathit{if}\ e_1 = \emptyset\ \mathit{then}\ e_2\ \mathit{else}\ e_3$, or *type switches* of the form $\mathit{if}\ e_1 \in \tau\ \mathit{then}\ e_2\ \mathit{else}\ e_3$. We denote the language of all positive-existential $\mathrm{RX}^-$ expressions by $\mathrm{PERX}^-$, and we will mention specific features added back to $\mathrm{PERX}^-$ in square brackets. Thus, $\mathrm{PERX}^-$[empty] includes emptiness tests, and $\mathrm{PERX}^-$[type] includes type switches.

**Proposition 1.** *$PERX^-[type]$ is equivalent to $RX^-$; in other words, type switches can be used to simulate emptiness tests.*

Indeed, *if* $e_1 = \emptyset$ *then* $e_2$ *else* $e_3$ can be expressed as follows:

$$\textit{if } (\textit{for } x \textit{ in } e_1 \textit{ return elem}\{a\}\{()\}) \in \mathbf{coll}(\mathbf{data}) \textit{ then } e_2 \textit{ else } e_3$$

The following proposition is not surprising, and parallels earlier results on semistructured query languages such as StruQL [10]:

**Proposition 2.** *$PERX^-[empty]$ can simulate the relational algebra. Concretely, for every relational algebra expression $\phi$ over database schema $\mathbf{S}$, there exists a $PERX^-[empty]$-expression $e_\phi$ and a type assignment $\Gamma_\mathbf{S}$, such that*

- *$e_\phi$ is well-defined under $\Gamma_\mathbf{S}$, and,*
- *$e_\phi$ evaluated on an encoding of database $D$ equals an encoding of $\phi(D)$.*

The simulation is described in Appendix A.

Consequently, satisfiability (i.e., nonempty output on at least one input) is undecidable for $\mathrm{PERX}^-$[empty] (and thus for $\mathrm{RX}^-$), because it is undecidable for the relational algebra. Since the expression

$$\textit{for } x \textit{ in } e \textit{ return elem}\{()\}\{()\}$$

is well-defined if, and only if, $e$ is unsatisfiable, we obtain:

**Corollary 1.** *Well-definedness for $PERX^-[empty]$ (and thus RX) is undecidable.*

What is perhaps more surprising is that without emptiness test, we remain undecidable:

**Theorem 1.** *Well-definedness for $PERX^-$ is undecidable.*

*Proof (Crux).* The proof goes by reduction from the implication problem for functional and inclusion dependencies, which is known to be undecidable [1, 8].

Let $\Sigma$ be a set of functional and inclusion dependencies, and let $\rho$ be an inclusion dependency. We show in Appendix B that we can construct two expressions $e_1$ and $e_2$, a type assignment $\Gamma$ and a node content type $\gamma$, such that

- $e_1$ and $e_2$ are well-defined under $\Gamma$,
- $\gamma$ is an output type for $e_1$ and $e_2$ under $\Gamma$, and,
- $e_1(\sigma) = e_2(\sigma)$ for every $\sigma \in \Gamma$ if, and only if, $\rho$ is implied by $\Sigma$.

Consequently, the expression $\textit{name}(\textit{elem}\{a\}\{e_1\}, \textit{elem}\{a\}\{e_2\})$ is well-defined under $\Gamma$ if, and only if, $\rho$ is implied by $\Sigma$. □

As a corollary to the proof, we note:

**Corollary 2.** *Equivalence of $PERX^-$ expressions is undecidable.*

We further derive:

**Corollary 3.** *Semantic type-checking for PERX$^-$ is undecidable.*

Indeed, referring to the above proof sketch of Theorem 1, $e_1$ and $e_2$ are equivalent if, and only if, $(\mathit{elem}\{a\}\{e_1\}, \mathit{elem}\{a\}\{e_2\})$ has output type $\mathbf{single}(\mathbf{elem}(\gamma))$.

We remark that to establish undecidability of well-definedness we do not need singleton types. For undecidability of semantic type-checking, we do.

## 5 Pure RX

In the XQuery data model, an item $i$ is identified with the singleton $\{i\}$ [11]. With this identification, it is indeed natural to let, e.g., $\mathit{name}(e)$ be undefined when $e(\sigma)$ is a set with more than one element. As we have seen in the previous Section, it is exactly this behavior that causes well-definedness to be undecidable.

So let us define a version of RX, called *pure RX*, which does not explicitly identify an item $i$ with $\{i\}$. We will show in Section 6 that well-definedness and semantic type-checking for the positive-existential fragment of pure RX is decidable.

A *pure RX-value* is an item or a set of items. A *pure RX-type* $\tau$ is a term generated by the following grammar:

$$\begin{aligned}\tau &::= \mathbf{coll}(\iota) \mid \iota \mid \tau \cup \tau \\ \iota &::= \mathbf{atom} \mid \nu \mid \iota \cup \iota \\ \nu &::= \mathbf{data} \mid \mathbf{elem}(\nu_1 \cup \cdots \cup \nu_k)\end{aligned}$$

Here, $\tau$ ranges over types, $\iota$ ranges over item types, $\nu$ ranges over node types, and $k \geq 0$.

A *pure RX-type* denotes a set of pure RX-values:

- **data** denotes the set of all data nodes,
- $\mathbf{elem}(\nu_1 \cup \cdots \cup \nu_k)$ denotes the set of all element nodes $\langle a : N \rangle$ for which $N$ is a finite set over the union of the denotations of $\nu_1, \ldots, \nu_k$.
- **atom** denotes the set $\mathcal{A}$ of all atoms,
- $\tau_1 \cup \tau_2$ denotes the union of the denotations of $\tau_1$ and $\tau_2$, and,
- $\mathbf{coll}(\iota)$ denotes the set of all finite sets over the denotation of $\iota$.

Note that since every $\iota$ is also a $\tau$, the denotation of $\iota_1 \cup \iota_2$ is subsumed by the definition above.

The syntax of *pure RX* is obtained from the syntax of RX by adding a singleton constructor expression $(e)$, and by replacing RX-types in type switch expressions by pure RX types.

In order to give the semantics of pure RX, we define the following (partial) functions on pure RX-values.

- $\mathit{data}'(v) = \{a \mid a \in v\} \cup \{a \mid \langle a \rangle \in v\}$

- $\mathit{name}'(v)$, which is $a$ if $v$ is an element node $\langle a : N\rangle$, and is undefined otherwise.
- $\mathit{children}'(v)$, which undefined if there is some atom in $v$, and otherwise returns

$$\bigcup\{N \mid \langle a : N\rangle \in v\}.$$

- $\mathit{construct}'(v, w)$ which is undefined if $v$ is not an atom, and returns $\langle v : N\rangle$ otherwise where $N$ is obtained from $w$ by replacing every atom in $w$ by a corresponding data node:

$$N = \{\langle a\rangle \mid a \in w\} \cup \{i \mid i \in w, i \text{ is a node}\}$$

The semantics of pure RX is then defined as follows:

$$\begin{aligned}
x(\sigma) &= \sigma(x) \\
\mathit{text}\{e\}(\sigma) &= \langle a\rangle \quad \text{if } e(\sigma) = a \\
\mathit{elem}\{e_1\}\{e_2\}(\sigma) &= \mathit{construct}'(e_1(\sigma), e_2(\sigma)) \\
\mathit{data}(e)(\sigma) &= \mathit{data}'(e(\sigma)) \\
\mathit{name}(e)(\sigma) &= \mathit{name}'(e(\sigma)) \\
\mathit{children}(e)(\sigma) &= \mathit{children}'(e(\sigma)) \\
()(\sigma) &= \emptyset \\
(e)(\sigma) &= \{e(\sigma)\} \quad \text{if } e(\sigma) \text{ is an item} \\
e_1, e_2(\sigma) &= e_1(\sigma) \cup e_2(\sigma) \\
\mathit{for}\ x : \kappa\ \mathit{in}\ e_1\ \mathit{return}\ e_2 &= \bigcup\{e_2(x : i, \sigma) \mid i \in e_1(\sigma) \cap \kappa\} \\
(\mathit{if}\ e_1\ \mathit{eq}\ e_2\ \mathit{then}\ e_3\ \mathit{else}\ e_4)(\sigma) &= \begin{cases} e_3(\sigma) & \text{if } e_1(\sigma), e_2(\sigma) \in \mathcal{A} \text{ and } e_1(\sigma) = e_2(\sigma) \\ e_4(\sigma) & \text{if } e_1(\sigma), e_2(\sigma) \in \mathcal{A} \text{ and } e_1(\sigma) \neq e_2(\sigma) \end{cases} \\
(\mathit{if}\ e_1 = \emptyset\ \mathit{then}\ e_2\ \mathit{else}\ e_3)(\sigma) &= \begin{cases} e_2(\sigma) & \text{if } e_1(\sigma) = \emptyset \\ e_3(\sigma) & \text{otherwise} \end{cases} \\
(\mathit{if}\ e_1 \in \tau\ \mathit{then}\ e_2\ \mathit{else}\ e_3)(\sigma) &= \begin{cases} e_2(\sigma) & \text{if } e_1(\sigma) \in \tau \\ e_3(\sigma) & \text{otherwise} \end{cases}
\end{aligned}$$

Note that again $e(\sigma)$ is not necessarily defined. Specifically, $e(\sigma)$ can become undefined for the following reasons:

- $e = \mathit{text}\{e'\}$, and $e'(\sigma)$ is not an atom,
- $e = \mathit{elem}\{e_1\}\{e_2\}$, and $e_1(\sigma)$ is not an atom,
- $e = \mathit{name}(e')$, and $e'(\sigma)$ is not an element node,
- $e = \mathit{children}(e')$, and $e'(\sigma)$ contains an atom,
- $e = (e')$, and $e'(\sigma)$ is not an item,
- $e = e_1, e_2$, and $e_1(\sigma)$ is not a set or $e_2(\sigma)$ is not a set, or,
- $e = \mathit{if}\ e_1\ \mathit{eq}\ e_2\ \mathit{then}\ e_3\ \mathit{else}\ e_4$, and $e_1(\sigma)$ or $e_2(\sigma)$ is not an atom.

**Pure PERX**

Clearly, well-definedness and semantic type-checking for the entire pure RX remains undecidable due to the presence of the emptiness test and type switch expressions. Let us define *pure PERX* as the fragment of pure RX in which these expressions are disallowed.

## 6 Decidability results

In this section we will show that well-definedness and semantic type-checking for pure PERX are decidable. In fact, we will solve the corresponding problems for the nested relational calculus (NRC): the well-known standard query language for nested relations and complex objects. Indeed, this language remains fundamental and its study remains interesting in its own right. As we will see, pure PERX can be simulated by the positive-existential fragment of NRC (extended with kind-tests).

### 6.1 Nested relational calculus

An *NRC-value* is either an atom, a pair of NRC-values, or a finite set of NRC-values. Note that we allow sets to be heterogeneous. If $v = (v_1, v_2)$, then we write $\pi_1(v)$ for $v_1$ and $\pi_2(v)$ for $v_2$.

An *NRC-type* $\tau$ is a term generated by the following grammar:

$$\tau ::= \emptyset \mid \mathbf{atom} \mid \tau \times \tau \mid \tau \cup \tau \mid \mathbf{coll}(\tau)$$

An NRC-type *denotes* a set of NRC-values:

- $\emptyset$ denotes the empty set,
- **atom** denotes the set $\mathcal{A}$ of all atoms,
- $\tau_1 \times \tau_2$ denotes the cartesian product of the denotations of $\tau_1$ and $\tau_2$,
- $\tau_1 \cup \tau_2$ denotes the union of the denotations of $\tau_1$ and $\tau_2$, and,
- $\mathbf{coll}(\tau)$ denotes the set of all finite sets over the denotation of $\tau$.

An *NRC-kind* $\kappa$ is a term generated by the following grammar:

$$\kappa ::= \mathbf{atom} \mid \mathbf{coll} \mid \kappa \times \kappa \mid \kappa \cup \kappa$$

An NRC-kind denotes a set of NRC-values, which can be the set of all atoms, the set of all finite sets of values, the cartesian product of the denotation of two kinds, or the union of the denotation of two kinds.

The *positive existential nested relational calculus* (PENRC) is the set of all expressions generated by the following grammar:

$$\begin{aligned} e ::= {} & x \\ \mid \ & (e, e) \mid \pi_1(e) \mid \pi_2(e) \\ \mid \ & \emptyset \mid \{e\} \mid e \cup e \mid \bigcup e \mid \{e \mid x \in e\} \\ \mid \ & e = e \mathbin{?} e : e \end{aligned}$$

Here $e$ ranges over expressions, and $x$ ranges over variables. The PENRC with kind tests, denoted by PENRC[kind] is the PENRC extended with one additional expression:

$$e ::= \cdots \mid e \in \kappa \ ?\ e \ :\ e$$

Here, $\kappa$ ranges over NRC kinds. The *free variables* of $e$ are defined in the usual way, and will be denoted by $FV(e)$.

If $e$ is a PENRC[kind]-expression and $\sigma$ is an NRC-environment on $e$, then the *semantics* $e(\sigma)$ of $e$ under $\sigma$ is inductively defined as follows:

$$\begin{aligned}
x(\sigma) &= \sigma(x) \\
(e_1, e_2)(\sigma) &= (e_1(\sigma), e_2(\sigma)) \\
\pi_1(e)(\sigma) &= \pi_1(e(\sigma)) \\
\pi_2(e)(\sigma) &= \pi_2(e(\sigma)) \\
\emptyset(\sigma) &= \emptyset \\
\{e\}(\sigma) &= \{e(\sigma)\} \\
(e_1 \cup e_2)(\sigma) &= e_1(\sigma) \cup e_2(\sigma) \\
(\bigcup e)(\sigma) &= \bigcup e(\sigma) \\
\{e_2 \mid x \in e_1\}(\sigma) &= \{e_2(x : v, \sigma) \mid v \in e_1(\sigma)\} \\
(e_1 = e_2 \ ?\ e_3 \ :\ e_4)(\sigma) &= \begin{cases} e_3(\sigma) & \text{if } e_1(\sigma), e_2(\sigma) \in \mathcal{A} \text{ and } e_1(\sigma) = e_2(\sigma) \\ e_4(\sigma) & \text{if } e_1(\sigma), e_2(\sigma) \in \mathcal{A} \text{ and } e_1(\sigma) \neq e_2(\sigma) \end{cases} \\
(e_1 \in \kappa \ ?\ e_2 \ :\ e_3)(\sigma) &= \begin{cases} e_2(\sigma) & \text{if } e_1(\sigma) \in \kappa \\ e_3(\sigma) & \text{otherwise} \end{cases}
\end{aligned}$$

Note that $e(\sigma)$ can be undefined. For example $\pi_1(x)(\sigma)$ is undefined when $\sigma(x)$ is not a pair, and $(x \cup y)(\sigma)$ is undefined when $\sigma(x)$ is not a set. Hence, we can also study the well-definedness problem for PENRC[kind].

It is easy to see that well-definedness for full NRC: PENRC extended with an emptiness test, is undecidable. Indeed, it is well known that NRC can simulate the relational algebra [6].

### 6.2 Simulating RX in NRC

Formally, a *simulation* of a query language $Q$ in a query language $Q'$ is a function $enc : V_Q \to V_{Q'}$ such that

- for every type $\tau \in T_Q$ there exits a type $\tau' \in T_{Q'}$ such that $v \in \tau$ if and only if $enc(v) \in \tau'$, and
- for every expression $e \in E_Q$ there exists an expression $e' \in E_{Q'}$ such that
  1. $e(\sigma)$ is defined if and only if $e'(enc(\sigma))$ is defined, and
  2. if $e(\sigma)$ is defined, then $enc(e(\sigma)) = e'(enc(\sigma))$.

A simulation is *effective* if $\tau'$ can be computed from $\tau$ and $e'$ can be computed from $e$.

**Lemma 1.** *Pure PERX can be effectively simulated in PENRC[kind].*

*Proof (Crux).* Consider the encoding function *enc* for which

$$enc(a) = a \qquad enc(\langle a\rangle) = ((a,a),\emptyset)$$
$$enc(\langle a : N\rangle) = (a, enc(N)) \qquad enc(v) = \{enc(i) \mid i \in v\}$$

Then *enc* is an effective simulation. It is easy to find $\tau'$ by induction on $\tau$. Furthermore, as shown in Appendix C, $e'$ can be constructed by induction on $e$. To illustrate this, we first introduce some syntactic sugar: we write $e_1 \in \kappa \to e_2$ for $e_1 \in \kappa \;?\; e_2 \;:\; \pi_1(\emptyset)$. Intuitively, this expression will be used to verify that the input to $e'$ is an encoding of a legal input to $e$. Otherwise, we become undefined.

We can now for example simulate $\langle e\rangle$ by $e' \in \mathbf{atom} \to ((e', e'), \emptyset)$. We can simulate $\langle e_1 : e_2\rangle$ by

$$e_1' \in \mathbf{atom} \to (e_1', \{x \in \mathbf{atom} \;?\; ((x,x),\emptyset) \;:\; x \mid x \in e_2'\}).$$

And we can simulate *children*$(e)$ by $\bigcup\{\pi_2(x) \mid x \in e'\}$. □

**Corollary 4.** *If the well-definedness or semantic type-checking problem is decidable for PENRC[kind], then it is also decidable for pure PERX.*

### 6.3 Well-definedness for pure PENRC[kind]

Consider the following expression:

$$e = \{\{z = y \;?\; \pi_1(z) \;:\; y \mid y \in x\}) \mid x \in R\}$$

and let the environment $\sigma$ be defined by

$$\sigma(R) = \{\{a,b\},\{c\},\{a,b,d\}\}\} \qquad \sigma(z) = d.$$

Since there is a set in $\sigma(R)$ which contains $\sigma(z)$, we will need to evaluate $\pi_1(\sigma(z))$ at some point, which is undefined. Hence, $e(\sigma)$ is undefined. Note that we do not need all elements in $\sigma(R)$ to reach the state where $e(\sigma)$ becomes undefined. Indeed, $e$ is also undefined on the small environment $\sigma'$ where $\sigma'(R) = \{\{d\}\}$ and $\sigma'(z) = d$.

We generalize this observation in the following general property, proven in Appendix D. Here, we say that an environment $\sigma$ is in the set $\mathcal{E}_k$ if every set occurring in $\sigma(x)$ has cardinality at most $k$ for every $x \in dom(\sigma)$.

**Lemma 2 (Small model property for undefinedness).** *Let $e$ be a PENRC[kind] expression, let $\Gamma$ be a type assignment on $e$, and let $\sigma$ be an environment compatible with $\Gamma$ such that $e(\sigma)$ is undefined. There exists a natural number $l$ which can be computed from $e$ alone, and an environment $\sigma' \in \mathcal{E}_l$ compatible with $\Gamma$, such that $e(\sigma')$ is also undefined.*

We obtain:

**Corollary 5.** *Well-definedness problem for PENRC[kind] is decidable.*

Indeed, up to isomorphism (and expressions cannot distinguish isomorphic inputs) there are only a finite number of different input environments in $\mathcal{E}_l$ compatible with $\Gamma$. So we can test them all to see if there is a counterexample to well-definedness.

Also for semantic type-checking we have:

**Lemma 3 (Small model property for semantic type-checking).** *Let $e$ be a PENRC[kind] expression, let $\Gamma$ be a type assignment on $e$ such that $e$ is well-defined under $\Gamma$, and let $\tau$ be a type. Let $\sigma$ be an environment compatible with $\Gamma$ such that $e(\sigma) \notin \tau$. There exists a natural number $l$ which can be computed from $e$ and $\tau$ alone, and an environment $\sigma' \in \mathcal{E}_l$ compatible with $\Gamma$, such that also $e(\sigma') \notin \tau$.*

**Corollary 6.** *Semantic type-checking for pure PENRC[kind] is decidable.*

### 6.4 Equivalence and satisfiability

The above decidability results are quite sharp, because *equivalence* of PENRC expressions is undecidable. This can be proven in a similar way as Theorem 1. Of course, containment is then also undecidable. Levy and Suciu [12] have shown that a "deep" form of containment (known as simulation) is decidable for PENRC.

Another important problem is satisfiability. For example, the XQuery type system generates a type error whenever it can deduce that an expression which is not the empty set expression itself always returns the empty set. As noted in Section 4, satisfiability is undecidable for $\text{PERX}^-$[empty]. For pure PERX, and PENRC[kind], however, satisfiability can be solved using the small model property for semantic type-checking. Indeed, a PENRC[kind] expression $e$ is satisfiable under $\Gamma$ if, and only if, $\mathbf{coll}(\emptyset)$ is an output type for $e$ under $\Gamma$. We point out that, at least for PENRC without union and kind-tests, decidability of satisfiability already follows from the work of Levy and Suciu cited above.

## 7 Lists and bags

In this paper we have focused our attention on a set-based abstraction of XQuery. The actual data model of XQuery is list-based however, and hence it is natural to ask how our results transfer to such a setting.

Let us denote by $\text{RX}^{\text{list}}$ the list-based version of RX, which can be obtained from RX as follows. The list-based RX data model is obtained by replacing "set" in the definition of the RX data model by "list". The list-based semantics of an expression is obtained from the set-based semantics by replacing every set operator by the corresponding list operator (i.e., empty set becomes empty list, union becomes concatenation, and so on). We can similarly define the bag-based version of RX, which we will denote by $\mathrm{RX}^{\mathrm{bag}}$.

We can still simulate the relational algebra in the list- and bag-based versions of $\mathrm{PERX}^{-}$[empty] and $\mathrm{PERX}^{-}$[type]. Hence, well-definedness and semantic type-checking for these languages is undecidable. It is an open problem however whether well-definedness and semantic type-checking in the list- and bag-based versions of $\mathrm{PERX}^{-}$ remains undecidable. Indeed, our undecidability proof depends heavily on the fact that set union is idempotent, which is not the case for list concatenation and bag union.

We can also consider a list-based and bag-based version of PENRC[kind], to which our decidability results transfer. Hence, well-definedness and semantic type-checking are decidable for pure $\mathrm{PERX}^{\mathrm{list}}$ and pure $\mathrm{PERX}^{\mathrm{bag}}$.

## A Relational Algebra Simulation

**Proposition 3.** *The relational algebra can be simulated in PERX$^-$[empty] under any interpretation of the "oracle" functions concat and content. Concretely, for every relational algebra expression $\phi$ over database schema* **S***, there exists a PERX$^-$[empty]-expression $e$ and a type assignment $\Gamma$, such that*

- *$e$ is well-defined under $\Gamma$, and,*
- *$e$ evaluated on an encoding of database $D$ equals an encoding of $\phi(D)$.*

*Proof.* In order to focus on the crux of the proof we will assume below that atoms are expressions in PERX$^-$[empty]. At the end of the proof we illustrate how we can get rid of this assumption.

There exists a well-known encoding of tuples as unordered trees. For example, the tuple $(A_1 : a_1, \dots, A_n : a_n)$ can be encoded as the element node

$$\langle T : \{\langle A_1 : \{a_1\}\rangle, \dots, \langle A_n : \{a_n\}\rangle\}\rangle.$$

Here we assume without loss of generality that attribute names are atoms. A relation can then be encoded as the set of the encodings of its tuples.

We assume without loss of generality that relation names are variables. Let **S** be a database schema, and let $\phi$ be a relational algebra expression over **S**. A database $D$ over **S** can be encoded as an environment $\sigma$ such that $\sigma(r)$ is an encoding of $D(r)$ for every relation name $r$.

We say that an expression $e$ with $FV(e) = FV(\phi)$ *simulates* $\phi$ if $e(\sigma)$ is an encoding of $\phi(D)$ whenever $\sigma$ is an encoding of $D$. We construct an expression $e_\phi$ which simulates $\phi$ by induction on $\phi$ as follows:[4]

- If $\phi = r$ then $e_\phi = r$.
- If $\phi = \sigma_{A_1=A_2}(\psi)$ then we define $e_\phi$ to be

```
for t in e_ψ return
 for x1, x2 in children(t) return
 if name(x1) = A1 and name(x2) = A2
 and children(x1) = children(x2) then
       t
   else
       ()
```

- If $\phi = \pi_{A_1,\dots,A_n}(\psi)$ then we define $e_\phi$ to be

```
for t in e_ψ return
   elem{T}{
     for x in children(t) return
```

[4] Here, we allow to bind multiple variables in one for loop, and also allow boolean combinations of conditions in an if test. Both features can clearly be simulated in PERX$^-$[empty].

```
    if name(x) = A_1 or ... or name(x) = A_n then
      x
    else
      ()
}
```

– If $\phi = \psi_1 \times \psi_2$ then we define $e_\phi$ to be (the schema's of $\psi_1$ and $\psi_2$ are disjoint):

```
for t_1 in e_ψ1, t_2 in e_ψ2 return
  elem{T}{children(t_1), children(t_2)}
```

– If $\phi = \rho_{A_1/A_2}(\psi)$ then we define $e_\phi$ to be

```
for t in e_ψ return
  elem{T}{
    for x in children(t) return
      if name(x) = A_1 then elem{A_2}{children(x)} else x
  }
```

– If $\phi = \psi_1 \cup \psi_2$ then we define $e_\phi$ as $e_{\phi_1}, e_{\phi_2}$.
– If $\phi = \psi_1 - \psi_2$ then we define $e_\phi$ as follows.

```
for t_1 in e_ψ1 return
 if f = ∅ then t_1 else ()
```

Here, the subexpression $f$ returns $\emptyset$ if $t_1 \notin e_{\psi_2}$, and $\{t_1\}$ otherwise. Let the schema of $\psi_1$ and $\psi_2$ be $\{A_1, \ldots, A_n\}$, then $f$ is constructed as follows:

```
for t_2 in e_ψ2 return
  for x_1,...,x_n in t_1 return
  for y_1,...,y_n in t_2 return
   if name(x_1) = A_1 and ... and name(x_n) = A_n
   and name(y_1) = A_1 and ... and name(y_n) = A_n
   and children(x_1) = children(y_1)
   and ...
   and children(x_n) = children(y_n) then
       t
   else
       ()
```

Note that $e_\phi$ is well-defined under every input $\sigma$ which is an encoding of some $D$. Furthermore, on such $\sigma$, there is never need to evaluate *concat* or *content* because:

– *name* is always evaluated on a singleton element node, and

– every if-test is between singleton data nodes (and hence, the application of *data* on the values which are to be compared can be calculated without using *content*).

Let $\Gamma$ be the type assignment for which for every $r \in FV(\phi)$

$$\Gamma(r) = \mathbf{coll}(\mathbf{elem}(\mathbf{coll}(\mathbf{elem}(\mathbf{single}(\mathbf{data}))))).$$

Note that $\sigma$ is not necessarily an encoding of a database $D$ when $\sigma$ is compatible with $\Gamma$. Hence, $e_\phi$ is not necessarily well-defined under $\Gamma$. However, we can transform every $\sigma(r)$ into an input which is an encoding of $D(r)$ for some database $D$ as follows. Let the schema $\mathbf{S}(r) = \{A_1, \ldots, A_n\}$.

```
for t in r return
 for x_1,...,x_n in t return
 if  name(x) = A_1 and ... and name(x_n) = A_n then
   elem{T}{x_1,...,x_n}
 else
   ()
```

Note that this expression does not change $\sigma(r)$ if it is already a valid encoding.

Let $e$ be the expression obtained by replacing every $r$ in $e_\phi$ by the expression above. It is easy to verify that now $e$ is well-defined under $\Gamma$.

*Note* We have assumed in the construction above that constant atoms (such as $A_1, \ldots, A_n$) are expressions. This makes the construction easier, but is not really necessary. We first note that we could have chosen any atom to represent an attribute $A_i$, as long as it differs from the atoms representing other attributes. Hence, we can create for every attribute name $A_i$ thus used a variable $x_{A_i}$ with type **single**(**atom**). An environment $\sigma$ is now an encoding of a database $D$ if (1) $\sigma(r)$ is an encoding of $D(r)$ (as before), and (2) $\sigma(x_{A_i}) \neq \sigma(x_{A_j})$ when $i \neq j$. Let $e'$ be the expression obtained from $e$ by replacing $A_i$ by $x_{A_i}$. Using if-tests we can then evaluate $e'$ if condition (2) holds, and evaluate () otherwise. □

## B Well-definedness in PERX$^-$

In this section we give the full proof that well-definedness for PERX$^-$ is undecidable.

**Theorem 2.** *Well-definedness for PERX$^-$ is undecidable.*

*Proof.* In order to focus on the crux of the proof we will assume below that atoms are expressions in PERX$^-$[empty]. At the end of the proof we illustrate how we can get rid of this assumption.

We give a reduction from the implication problem for functional and inclusion dependencies, which is known to be undecidable [1, 8]. We assume without loss of generality that the database schema consists of a single relation symbol with schema $\{A_1, \ldots, A_k\}$. A functional dependency $\phi$ is a rule $X \to Y$ where $X$ and $Y$ are subsets of $\{A_1, \ldots, A_k\}$. We say that relation $R$ satisfies $\phi$, denoted by $R \models \phi$ if for all tuples $t_1, t_2 \in R$ if $\pi_X(t_1) = \pi_X(t_2)$ then also $\pi_Y(t_1) = \pi_Y(t_2)$. An inclusion dependency $\psi$ is a rule of the form $[B_1, \ldots, B_i] \to [C_1, \ldots, C_i]$ where $\{B_1, \ldots, B_i\}$ and $\{C_1, \ldots, C_i\}$ are subsets of $\{A_1, \ldots, A_k\}$. We say that relation $R$ satisfies $\psi$, denoted by $R \models \psi$ if $\pi_{B_1,\ldots,B_k}(R) \subseteq \pi_{C_1,\ldots,C_k}(R)$.

Let $\Sigma$ be a set of functional and inclusion dependencies. Let $\phi_1, \ldots, \phi_n$ be the functional dependencies in $\Sigma$, and let $\psi_1, \ldots, \psi_m$ be the inclusion dependencies in $\Sigma$. Let $\rho$ be an additional target functional dependency. The implication problem consists of checking for every relation $R$ over $\{A_1, \ldots, A_k\}$ that if $R$ satisfies every dependency in $\Sigma$, then $R$ also satisfies $\rho$.

We assume without loss of generality that $A_1, \ldots, A_k$ are atoms. We will encode a $k$-tuple $(A_1 : a_1, \ldots, A_k : a_k)$ by the element node

$$\langle A_1 : \{\langle A_1 : \{a_1\}\rangle, \ldots, \langle A_k : \{a_k\}\rangle\}\rangle.$$

A relation $R$ is then encoded as the set of the encodings of tuples in $R$. Let $r$ be a variable, and let $\Gamma$ be the type assignment for which

$$\Gamma(r) = \mathbf{coll}(\mathbf{elem}(\mathbf{coll}(\mathbf{elem}(\mathbf{single}(\mathbf{data}))))).$$

Note that $\sigma(r)$ is not necessarily an encoding of some relation $R$ when $\sigma$ is compatible with $\Gamma$. For example, suppose that $k \geq 2$, and consider

$$\sigma(r) = \{\langle A_1 : \{\langle A_1 : \{\langle a\rangle\}\rangle\}\rangle\}.$$

Then $\sigma \in \Gamma$, but there is clearly no instance of $R$ for which $\sigma(r)$ is an encoding.

However, we can transform every input into an input which is an encoding of some relation:[5]

```
for t1 in r return
 for x1,...,xk in t1 return
 if  name(x1) = A1 and ... and name(xk) = Ak then
   elem{A1}{x1,...,xk}
 else ()
```

The output of this expression is always a set of element nodes which are encodings of $k$-ary tuples. In the remainder of this proof we will therefore assume w.l.o.g. that $r$ contains an actual encoding of some relation $r$: we can always replace $r$ by the expression above.

For every functional dependency $\phi = \{B_1, \ldots, B_i\} \to \{C_1, \ldots, C_j\}$, we create the following expression $e_\phi$:

```
for t1,t2 in r return
 for x1,...,xi,y1,...,yj in t1 return
```

[5] Here, we allow to bind multiple variable in one for loop, and also allow boolean combinations of conditions in an if test. Both features can clearly be simulated in PERX$^-$.

```
for u_1, ..., u_i, v_1, ..., v_j in t_2 return
 if  name(x_1) = B_1 and ... and name(x_i) = B_i
 and name(y_1) = C_1 and ... and name(y_j) = C_j
 and name(u_1) = B_1 and ... and name(u_i) = B_i
 and name(v_1) = C_1 and ... and name(v_j) = C_j then
  if  children(x_1) = children(u_1)
  and ...
  and children(x_i) = children(u_i)
  and (children(y_1)  ≠  children(v_1)
       or ...
       or children(y_j)  ≠  children(v_j)
      ) then elem{A_1}{()} else ()
 else ()
```

Note that on input $\sigma$ for which $\sigma(r)$ is an encoding of relation $R$ this expression returns $\emptyset$ if $R \models \phi$, and $\langle A_1 : \emptyset\rangle$ otherwise.

For every inclusion dependency $\psi = R : [B_1, \ldots, B_i] \subseteq S : [C_1, \ldots, C_i]$ we create the following expression $e_\psi$:

```
for t_1 in r return elem{A_1}{
 for t_2 in r return
  for x_1, ..., x_i in t_1 return
  for y_1, ..., y_i in t_2 return
   if  name(x_1) = B_1 and ... and name(x_i) = B_i
   and name(y_1) = C_1 and ... and name(y_i) = C_i
   and  children(x_1) = children(y_1)
   and ...
   and children(x_i) = children(y_i)
     then elem{A_1}{()} else ()
}, elem{A_1}{elem{A_1}{()}}
```

Note that on input $\sigma$ for which $\sigma(r)$ is an encoding of relation $R$, this expression returns $\{\langle A_1 : \{\langle A_1 : \emptyset\rangle\}\rangle\}$ if $R \models \phi$, and $\{\langle A_1 : \{\langle A_1 : \emptyset\rangle\}\rangle, \langle A_1 : \emptyset\rangle\}$ otherwise.

Let $D_0, \ldots, D_n, E_1, \ldots, E_m$ be different elements of $\mathcal{A}$. We then create the expression $e_1$ as follows:

```
elem{A_1}{
  elem {D_0}{e_ρ},elem {D_1}{e_φ1},...,elem {D_n}{e_φn},
  elem {E_1}{e_ψ1}, ..., elem {E_m}{e_ψm}
}
```

Let $f_1, \ldots, f_n, h$ be expressions of the form () or *elem*$\{A_1\}\{()\}\}$, and let $g_1, \ldots, g_m$ be expressions of the form (*elem*$\{A_1\}\{()\}$, *elem*$\{A_1\}\{$*elem*$\{A_1\}\{()\}\}$), or *elem*$\{A_1\}\{$*elem*$\{A_1\}\{()\}\}$. We call an expression

```
elem{A_1}{
  elem {D_0}{h},elem {D_1}{f_1},...,elem {D_n}{f_n},
  elem {E_1}{g_1}, ..., elem {E_m}{g_m}
}
```

*admissible* when, if $f_1 = \cdots = f_n = ()$ and $g_1 \ldots g_m = \mathit{elem}\{A_1\}\{\mathit{elem}\{A_1\}\{()\}\}$, then $h = ()$. Clearly the number of admissible expressions is finite. Let $\{i_1, \ldots, i_k\}$ be the set of all admissible expressions. Then we define $e_2$ as the expression $i_1, \ldots, i_n$.

Note that $e_1$ and $e_2$ are well-defined under $\Gamma$. Furthermore, the following node content type is an output type for $e_1$ and $e_2$ under $\Gamma$:

$$\gamma := \mathbf{coll}(\mathbf{elem}(\mathbf{coll}(\mathbf{elem}(\mathbf{coll}(\mathbf{elem}(\mathbf{coll}(\mathbf{elem}(\mathbf{coll}(\mathbf{data}))))))))).$$

Also, $\Sigma \models \rho$ if, and only if, $e_1(\sigma) \subseteq e_2(\sigma)$ for every $\sigma \in \Gamma$. Hence, $\Sigma \models \rho$ if, and only if, $(e_1, e_2)(\sigma) = e_2(\sigma)$ for every $\sigma \in \Gamma$. Consequently,

$$e := \mathit{name}(\mathit{elem}\{A_1\}\{e_1 \cup e_2\}, \mathit{elem}\{A_1\}\{e_2\})$$

is defined if and only if $\Sigma \models \rho$. Furthermore, $\mathit{elem}\{A_1\}\{e_1 \cup e_2\}, \mathit{elem}\{A_1\}\{e_2\})$ has output type $\mathbf{single}(\mathbf{elem}(\gamma))$ if, and only if $\Sigma \models \rho$.

*Note* We have assumed in the construction above that constant atoms (such as $A_1, \ldots, A_n$) are expressions. This makes the construction easier, but is not really necessary. We first note that we could have chosen any atom to represent an attribute $A_i$, as long as it differs from the atoms representing other attributes. Hence, we can create for every attribute name $A_i$ thus used a variable $x_{A_i}$ with type $\mathbf{single}(\mathbf{atom})$. An environment $\sigma$ is now an encoding of a relation $R$ if (1) $\sigma(r)$ is an encoding of $R$ (as before), and (2) $\sigma(x_A) \neq \sigma(x_B)$ when $A \neq B$. Let $e'$ be the expression obtained from $e$ by replacing every $A$ by $x_A$. Using if-tests we can then evaluate $e'$ if condition (2) holds, and evaluate $()$ otherwise. □

## C Simulating pure PERX in PENRC[kind]

In this section we give the full proof of the simulation of PERX in PENRC[kind].

**Lemma 4.** *The pure PERX can effectively be simulated in PENRC[kind].*

*Proof.*

$$\mathit{enc}(a) = a \qquad \mathit{enc}(\langle a \rangle) = ((a, a), \emptyset)$$
$$\mathit{enc}(\langle a : N \rangle) = (a, \mathit{enc}(N)) \qquad \mathit{enc}(v) = \{\mathit{enc}(i) \mid i \in v\}$$

We will now verify that *enc* has the required properties. Let $\tau$ be a pure RX-type. We define the pure NRC-type $\tau'$ by induction on $\tau$ as follows.

$$\begin{aligned}
\mathbf{atom}' &= \mathbf{atom} \\
\mathbf{data}' &= (\mathbf{atom} \times \mathbf{atom}) \times \mathbf{coll}(\emptyset) \\
\mathbf{elem}(\nu_1 \cup \cdots \cup \nu_k)' &= \mathbf{atom} \times \mathbf{coll}(\nu_1' \cup \cdots \cup \nu_k') \\
\mathbf{coll}(\iota)' &= \mathbf{coll}(\iota') \\
(\tau_1 \cup \tau_2)' &= \tau_1' \cup \tau_2'
\end{aligned}$$

Obviously, $v \in \tau$ if, and only if $enc(v) \in \tau'$. If $\kappa$ is a pure PERX-kind, then we define the PENRC-kind $\kappa'$ by induction on $\kappa$ as follows:

$$\begin{aligned}
\mathbf{atom}' &= \mathbf{atom} \\
\mathbf{data}' &= (\mathbf{atom} \times \mathbf{atom}) \times \mathbf{coll} \\
\mathbf{elem}' &= \mathbf{atom} \times \mathbf{coll} \\
(\kappa_1 \cup \kappa_2)' &= \kappa_1' \cup \kappa_2'
\end{aligned}$$

Obviously, $v \in \kappa$ if, and only if, $enc(v) \in \kappa'$.

Let $e$ be a pure PERX-expression. We will define the pure PENRC-expression $e'$ by induction on $e$ as follows. We first introduce some syntactic sugar: we write $e_1 \in \kappa \rightarrow e_2$ for $e_1 \in \kappa \ ?\ e_2 \ :\ \pi_1(\emptyset)$. Intuitively, this expression will be used to verify that the input to $e'$ is an encoding of a legal input to $e$. Otherwise, we become undefined.

$$\begin{aligned}
x' &:= x \\
\langle e \rangle' &:= e' \in \mathbf{atom} \rightarrow ((e', e'), \emptyset) \\
\langle e_1 : e_2 \rangle' &:= e_1' \in \mathbf{atom} \rightarrow \\
&\qquad (e_1', \{x \in \mathbf{atom} \ ?\ ((x, x), \emptyset) \ :\ x \mid x \in e_2'\}) \\
\mathit{data}(e)' &:= \bigcup \{x \in (\mathbf{atom} \times \mathbf{atom}) \times \mathbf{coll} \ ?\ \{\pi_1(\pi_1(x))\} \\
&\qquad : x \in \mathbf{atom} \ ?\ \{x\} \ :\ \emptyset \mid x \in e'\} \\
\mathit{name}(e)' &:= \pi_1(e') \in \mathbf{atom} \rightarrow \pi_1(e') \\
\mathit{children}(e)' &:= \bigcup \{\pi_2(x) \mid x \in e'\} \\
()' &:= \emptyset \\
(e)' &:= e' \in \mathbf{atom} \cup ((\mathbf{atom} \times \mathbf{atom}) \times \mathbf{coll}) \cup (\mathbf{atom} \times \mathbf{coll}) \rightarrow \{e'\} \\
(e_1, e_2)' &:= e_1' \cup e_2' \\
(\mathit{for}\ x : \kappa\ \mathit{in}\ e_1\ \mathit{return}\ e_2)' &:= \bigcup \{x \in \kappa' \ ?\ e_2' \ :\ \emptyset \mid x \in e_1'\} \\
(\mathit{if}\ e_1\ \mathit{eq}\ e_2\ \mathit{then}\ e_3\ \mathit{else}\ e_4)' &:= e_1' = e_2' \ ?\ e_3' \ :\ e_4'
\end{aligned}$$

A straightforward induction on $e$ now shows that

1. $e(\sigma)$ is defined if, and only if $e'(enc(\sigma))$ is defined, and
2. if $e(\sigma)$ is defined then $enc(e(\sigma)) = e'(enc(\sigma))$. □

## D Well-definedness for PENRC[kind]

In this section we give the proof of the small model properties mentioned in Section 6.3. Whenever we write "expression" we mean PENRC[kind]-expression, and whenever we write type we mean PENRC-type.

Let $k$ be a natural number. We write $\mathcal{V}_k$ for the set of all values $v$ for which the cardinality of a set occurring in $v$ is at most $k$:

$$\mathcal{V}_k = \mathcal{A} \cup (\mathcal{V}_k \times \mathcal{V}_k) \cup \{v \subseteq \mathcal{V}_k \mid |v| \leq k\}.$$

We write $\mathcal{E}_k$ for the set of all environments $\sigma$ for which the cardinality of a set occurring in $\sigma$ is at most $k$:

$$\mathcal{E}_k = \{\sigma \mid \sigma \text{ environment on some } X \text{ and } \sigma(x) \in \mathcal{V}_k \text{ for all } x \in X\}.$$

The *sub-value* relation $\sqsubseteq$ on values is inductively defined by the following inference rules:

$$\frac{}{a \sqsubseteq a} \qquad \frac{v \sqsubseteq v' \qquad w \sqsubseteq w'}{(v,w) \sqsubseteq (v',w')} \qquad \frac{\text{for all } v_i \text{ there exists } w_j \text{ such that } v_i \sqsubseteq w_j}{\{v_1,\ldots,v_n\} \sqsubseteq \{w_1,\ldots,w_m\}}$$

This relation can be extended component-wise to environments: let $\sigma$ and $\sigma'$ be two environments on the same set of variables $X$, then $\sigma \sqsubseteq \sigma'$ if $\sigma(x) \sqsubseteq \sigma'(x)$ for all $x \in X$.

If $u, v, w$ are values, and $u \sqsubseteq w$ and $v \sqsubseteq w$, then we define $u \sqcup v$ as follows:

$$a \sqcup a = a \qquad (u_1,u_2) \sqcup (w_1,w_2) = (u_1 \sqcup w_1, u_2 \sqcup w_2) \qquad v \sqcup w = v \cup w$$

**Lemma 5.** *If $u \sqsubseteq w$ and $v \sqsubseteq w$, then $u \sqsubseteq u \sqcup v$, $v \sqsubseteq u \sqcup v$, and $u \sqcup v \sqsubseteq w$. Moreover, if $u \in \mathcal{V}_k$ and $v \in \mathcal{V}_l$, then $u \sqcup v \in \mathcal{V}_{k+l}$.*

*Proof.* By a straightforward induction on $w$.

We first give some characteristics of $\sqsubseteq$ with regard to kinds and types.

**Lemma 6.** *If $v \sqsubseteq w$ then $v \in \kappa$ if, and only if, $w \in \kappa$.*

*Proof.* By a trivial induction on $\kappa$.

**Lemma 7.** *Let $\tau$ be a type. If $v \sqsubseteq w$ and $w \in \tau$, then $v \in \tau$.*

*Proof.* By a trivial induction on $\tau$.

The PENRC has the following monotonicity properties. The proof is by a straightforward induction.

**Lemma 8 (Monotonicity).** *Let $e$ be an expression, and let $\sigma$ and $\sigma'$ be environments on $FV(e)$ such that $\sigma \sqsubseteq \sigma'$. If $e(\sigma)$ and $e(\sigma')$ are defined, then $e(\sigma) \sqsubseteq e(\sigma')$. If $e(\sigma)$ is undefined, then so is $e(\sigma')$.*

In what follows we will frequently take the *minimum* $\min(v)$ of a value $v$, which is obtained by replacing every set occurring in $v$ by the empty set. For example, $\min((\{a,b\},(a,\{\{c,d\}\}))) = (\emptyset,(a,\emptyset))$. Obviously, $\min(v) \in \mathcal{V}_0$ and $\min(v) \sqsubseteq v$.

Before we are ready to state our small model properties we need one final definition. Let $e$ be an expression, and let $k$ be a natural number. The *$k$-complexity* $c(e,k)$ of $e$ is inductively defined as follows:

$$
\begin{aligned}
&c(x,k) = k\\
&c((e_1,e_2),k) = c(e_1 \cup e_2, k) = c(e_1,k) + c(e_2,k)\\
&c(\pi_1(e'),k) = c(\pi_2(e'),k) = c(\bigcup e') = c(e',k)\\
&c(\{e'\},k) = k \times c(e',k)\\
&c(\emptyset,k) = 0\\
&c(\{e_2 \mid x \in e_1\},k) = c(e_1, \max(k, c(e_2,k))) + k \times c(e_2,k)\\
&c(e_1 = e_2 \; ? \; e_3 \; : \; e_4, k) = \max(c(e_3,k), c(e_4,k))\\
&c(e_1 \in \kappa \; ? \; e_2 \; : \; e_3, k) = \max(c(e_2,k), c(e_3,k))
\end{aligned}
$$

**Lemma 9 (Small model).** *Let $k$ be a natural number, let $e$ be an expression, and let $\sigma$ be an environment on $FV(e)$ such that $e(\sigma)$ is defined. Let $u \in \mathcal{V}_k$. If $u \sqsubseteq e(\sigma)$, then there exists an environment $\sigma' \in \mathcal{E}_{c(e,k)}$ with $\sigma' \sqsubseteq \sigma$ such that $u \sqsubseteq e(\sigma')$.*

*Proof.* Note that since $e(\sigma)$ is defined, $e(\delta)$ is also defined for every $\delta \sqsubseteq \sigma$ by monotonicity. Moreover, if $\sigma_1 \sqsubseteq \sigma$ and $\sigma_2 \sqsubseteq \sigma$, then $\sigma_1 \sqsubseteq \sigma_1 \sqcup \sigma_2$, $\sigma_2 \sqsubseteq \sigma_1 \sqcup \sigma_2$, and $\sigma_1 \sqcup \sigma_2 \sqsubseteq \sigma$ by Lemma 5. We will use these facts silently throughout this proof.

In order to write the proof in a succinct manner, let us define the set $P(u,e,\sigma,k)$ by

$$P(u,e,\sigma,k) = \{\sigma' \mid \sigma' \in \mathcal{E}_{c(e,k)}, \sigma' \sqsubseteq \sigma, \text{ and } u \sqsubseteq e(\sigma')\}.$$

We will prove by induction on $e$ that $P(u,e,\sigma,k)$ is non-empty.

– If $e = x$, then we define $\sigma'$ by

$$\sigma'(y) = \begin{cases} u & \text{if } y = x\\ \min(\sigma(y)) & \text{otherwise} \end{cases}$$

– If $e = \emptyset$, then we take $\sigma' = \min(\sigma)$.
– If $e = (e_1,e_2)$, then $e(\sigma)$ is a pair. Hence, $u = (u_1,u_2)$ for some $u_1,u_2 \in \mathcal{V}_k$. By the induction hypothesis there exist $\sigma_1 \in P(u_1,e_1,\sigma,k)$ and $\sigma_2 \in P(u_2,e_2,\sigma,k)$. Then $\sigma_1 \sqcup \sigma_2 \in \mathcal{E}_{c(e_1,k)+c(e_2,k)} = \mathcal{E}_{c(e,k)}$. Then, by monotonicity:

$$(u_1,u_2) \sqsubseteq (e_1(\sigma_1), e_2(\sigma_2)) \sqsubseteq (e_1(\sigma_1 \sqcup \sigma_2), e_2(\sigma_1 \sqcup \sigma_2)) = e(\sigma_1 \sqcup \sigma_2)$$

Hence, $\sigma_1 \sqcup \sigma_2 \in P(u,e,\sigma,k)$.

– If $e = e_1 \cup e_2$, then $e(\sigma)$ is a set. Since $u \sqsubseteq e(\sigma)$ there exists $w_v \in e(\sigma)$ with $v \sqsubseteq w_v$ for every $v \in u$. Define

$$u_1 = \{v \in u \mid w_v \in e_1(\sigma)\}$$
$$u_2 = \{v \in u \mid w_v \in e_2(\sigma)\}$$

Then $u = u_1 \cup u_2$, $u_1 \sqsubseteq e_1(\sigma)$, and $u_2 \sqsubseteq e_2(\sigma)$. Moreover, $u_1, u_2 \in \mathcal{V}_k$. The result then follows from the induction hypothesis by a reasoning similar to the previous case.

– If $e = \pi_1(e')$, then $e'(\sigma)$ is a pair $(v, w)$. Let $u' = (u, \min(w))$. Then $u' \sqsubseteq (v, w)$ since $u \sqsubseteq v$ and $\min(w) \sqsubseteq w$. Moreover, $u' \in \mathcal{V}_k$ since $\min(w) \in \mathcal{V}_0$. Hence there exists $\sigma' \in P(u', e', \sigma, k)$ by the induction hypothesis. Hence, $u = \pi_1(u') \sqsubseteq \pi_1(e'(\sigma')) = e(\sigma')$. Since $\mathcal{E}_{c(e',k)} = \mathcal{E}_{c(e,k)}$, $\sigma' \in P(u, e, \sigma, k)$. The case where $e = \pi_2(e')$ is similar.

– If $e = \{e'\}$ then we discern two cases. If $u = \emptyset$, then $u \sqsubseteq e(\sigma')$ for any $\sigma' \sqsubseteq \sigma$ by monotonicity. Hence, it suffices to take $\sigma' = \min(\sigma)$, which is in $\mathcal{E}_0$. Otherwise, $u$ contains at least one and at most $k$ elements. Let $v \in u$. Then $v \in \mathcal{V}_k$, and $v \sqsubseteq e'(\sigma)$ since $u \sqsubseteq e(\sigma)$. By the induction hypothesis there exists $\sigma_v \in P(v, e', \sigma, k)$. Let $\sigma' = \bigsqcup_{v \in u} \sigma_v$. Then $\sigma_v \sqsubseteq \sigma'$ for every $v \in u$, and $\sigma' \sqsubseteq \sigma$. By monotonicity we then have

$$v \sqsubseteq e'(\sigma_v) \sqsubseteq e'(\sigma').$$

And hence $u \sqsubseteq e'(\sigma')$. Moreover, $\sigma' \in \mathcal{E}_{k \times c(e',k)} = c(e, k)$ by Lemma 5. Hence, $\sigma' \in P(u, e, \sigma, k)$.

– If $e = \bigcup e'$ then $e'(\sigma)$ is a set of sets. For every $v \in u$ there exists $w_v \in e(\sigma)$ such that $v \sqsubseteq w_v$ since $u \sqsubseteq e(\sigma)$.
Let $e'(\sigma) = \{z_1, \dots, z_n\}$. Define

$$u_i = \{v \in u \mid w_v \in z_i \setminus \bigcup_{j<i} z_j\}.$$

Note that the cardinality of each of the $u_i$'s is at most $k$, and that at most $k$ of the $u_i$'s are non-empty. Furthermore, $u_i \sqsubseteq z_i$. Let $u'$ be the set of all non-empty $u_i$'s. Then $u' \sqsubseteq e'(\sigma)$ and $u' \in \mathcal{V}_k$. The result then follows from the induction hypothesis.

– If $e = e_1 = e_2 \ ?\ e_3 \ :\ e_4$, then $e_1(\sigma), e_2(\sigma) \in \mathcal{A}$. Suppose $e_1(\sigma) = e_2(\sigma)$, then $u \sqsubseteq e_3(\sigma)$. By the induction hypothesis there exists $\sigma' \in P(u, e_3, \sigma, k)$. Then $e_1(\sigma') = e_1(\sigma) = e_2(\sigma) = e_2(\sigma')$ by monotonicity and hence $e(\sigma') = e_3(\sigma')$. We then have by the induction hypothesis that $u \sqsubseteq e_3(\sigma') = e(\sigma')$. Since $\sigma' \in \mathcal{E}_{c(e_3,k)} \subseteq \mathcal{E}_{c(e,k)}$, $\sigma' \in P(u, e, \sigma, k)$. The case where $e_1(\sigma) \neq e_2(\sigma)$ is similar.

– If $e = e_1 \in \kappa \ ?\ e_2 \ :\ e_3$, we discern two cases. If $e_1(\sigma) \in \kappa$ then $u \sqsubseteq e_2(\sigma)$. By the induction hypothesis there exist $\sigma' \in P(u, e_2, \sigma, k)$. By monotonicity, $e_1(\sigma') \sqsubseteq e_1(\sigma)$. Hence, $e_1(\sigma') \in \kappa$ by Lemma 6. Then $e(\sigma') = e_2(\sigma')$, and hence $u \sqsubseteq e_2(\sigma') = e(\sigma')$. Then $\sigma' \in P(u, e, \sigma, k)$ since $\sigma' \in \mathcal{E}_{c(e_2,k)} \subseteq \mathcal{E}_{c(e,k)}$. The case where $e_1(\sigma) \notin \kappa$ is similar.

– If $e = \{e_2 \mid x \in e_1\}$, then $e(\sigma)$ is a set. Let $v \in u$. Since $u \sqsubseteq e(\sigma)$ there exists $w_v \in e(\sigma)$ such that $v \sqsubseteq w_v$. Since $e(\sigma)$ is obtained by a comprehension over $e_1(\sigma)$, there also must exists some $z_v \in e_1(\sigma)$ such that $v \sqsubseteq w_v = e_2(x : z_v, \sigma)$. Hence, there exists $x : z'_v, \sigma'_v \in P(v, e_2, (x : z_v, \sigma), k)$ by the induction hypothesis. Let $u' = \{z'_v \mid v \in u\}$. Then $u'$ contains at most $k$ elements of $\mathcal{V}_{c(e_2,k)}$. Hence, $u' \in \mathcal{V}_{\max(k,c(e_2,k))}$. Moreover, $x : z'_v, \sigma' \sqsubseteq x : z_v, \sigma$ by the induction hypothesis, so $z'_v \sqsubseteq z_v$, and hence $u' \sqsubseteq e_1(\sigma)$.
  By applying the induction hypothesis again, there exists $\sigma_1 \in P(u', e_1, \sigma, k)$. Let $\sigma' = \sigma_1 \sqcup \bigsqcup_{v \in u} \sigma'_v$. Note that $\sigma_1 \sqsubseteq \sigma'$, and $\sigma'_v \sqsubseteq \sigma'$ for every $v \in u$. Furthermore, $\sigma' \sqsubseteq \sigma$ and the maximum cardinality of a set in $\sigma'$ is bounded by (Lemma 5):

$$c(e_1, \max(k, c(e_2, k))) + k \times c(e_2, k) = c(e, k)$$

  Now $u' \sqsubseteq e_1(\sigma_1) \sqsubseteq e_1(\sigma')$ by monotonicity. Hence, for every $z'_v$ there exists some $z''_v \in e_1(\sigma')$ with $z'_v \sqsubseteq z''_v$. Then $x : z'_v, \sigma'_v \sqsubseteq x : z''_v, \sigma'$, and hence $v \sqsubseteq e_2(x : z'_v, \sigma'_v) \sqsubseteq e_2(x : z''_v, \sigma')$ by monotonicity. Since this holds for every $v \in u$, we have $u \sqsubseteq e(\sigma')$. □

**Lemma 10.** *Let $e$ be an expression and $\sigma$ an environment on $FV(e)$. If $e(\sigma)$ is undefined, then there exists $\sigma' \in \mathcal{E}_{c(e,1)}$ with $\sigma' \sqsubseteq \sigma$ such that $e(\sigma')$ is undefined.*

*Proof.* The proof is by induction on $e$. We note again that if $e'(\sigma)$ is defined, then $e'(\delta)$ is also defined for every $\delta \sqsubseteq \sigma$ by monotonicity.

– If $e = x$, or $e = \emptyset$, then there is nothing to prove, since $e(\sigma)$ is always defined.
– If $e = (e_1, e_2)$, then either $e_1(\sigma)$ or $e_2(\sigma)$ is undefined. The result then follows by the induction hypothesis. The case where $e = \{e'\}$ is similar.
– If $e = \pi_1(e')$, then either $e'(\sigma)$ is undefined, in which case the result follows from the induction hypothesis, or $e'(\sigma)$ is not a pair. Let $\sigma' = \min(\sigma)$. By monotonicity $e'(\sigma')$ cannot be a pair, and hence $e(\sigma')$ is also undefined. Moreover, $\sigma' \in \mathcal{E}_0 \subseteq \mathcal{E}_{c(e,1)}$.
– If $e = e_1 \cup e_2$, then either $e_1(\sigma)$ is undefined, $e_2(\sigma)$ is undefined, $e_1(\sigma)$ is not a set, or $e_2(\sigma)$ is not a set. In the first two cases the result follows from the induction hypothesis. In the third cases, let $\sigma' = \min(\sigma)$. By monotonicity $e_1(\sigma)$ cannot be a set, and hence $e(\sigma')$ is undefined. Moreover, $\sigma' \in \mathcal{E}_0 \subseteq \mathcal{E}_{c(e,1)}$. The last case is similar.
– If $e = \bigcup e'$, then either $e'(\sigma)$ is undefined, or $e'(\sigma)$ is not a set of sets. In the first case the result follows from the induction hypothesis. In the latter case we have two possibilities. If $e'(\sigma)$ is not a set, then let $\sigma' = \min(\sigma)$. By monotonicity, $e'(\sigma')$ cannot be a set, and hence $e(\sigma)$ is undefined. Moreover, $\sigma' \in \mathcal{E}_0 \subseteq \mathcal{E}_{c(e,1)}$. If $e'(\sigma)$ is a set, but not a set of sets, then there exist some $u \in e'(\sigma)$ that is not a set. Then $\{\min(u)\} \in \mathcal{V}_1$, and $\{\min(u)\} \sqsubseteq e'(\sigma)$. By Lemma 9 there exists $\sigma'' \in \mathcal{E}_{c(e',1)} = \mathcal{E}_{c(e,1)}$ with $\sigma'' \sqsubseteq \sigma$ such that $\{\min(u)\} \sqsubseteq e'(\sigma'')$. Hence, $e'(\sigma)$ is not a set of sets, and $e(\sigma'')$ is also undefined.

- If $e = e_1 = e_2 \mathrel{?} e_3 : e_4$, then we have three possibilities. If $e_1(\sigma)$ or $e_2(\sigma)$ is undefined, then the result follows from the induction hypothesis. If $e_1(\sigma)$ and $e_2(\sigma)$ are defined and $e_1(\sigma) = e_2(\sigma)$, then $e_3(\sigma)$ must be undefined. By the induction hypothesis, there exists $\sigma' \in \mathcal{E}_{c(e_3,1)} \subseteq \mathcal{E}_{c(e,1)}$ with $\sigma' \sqsubseteq \sigma$ such that $e_3(\sigma')$ is also undefined. By monotonicity $e_1(\sigma') = e_2(\sigma')$, and hence $e(\sigma')$ is undefined. If $e_1(\sigma) \neq e_2(\sigma)$ the reasoning is similar.
- If $e = e_1 \in \kappa \mathrel{?} e_2 : e_3$, then we have three possibilities. If $e_1(\sigma)$ is undefined, then the result follows from the induction hypothesis. If $e_1(\sigma)$ is defined and $e_1(\sigma) \in \kappa$, then $e_2(\sigma)$ must be undefined. By the induction hypothesis we have $\sigma' \in \mathcal{E}_{c(e_2,1)} \subseteq \mathcal{E}_{c(e,1)}$ with $\sigma' \sqsubseteq \sigma$ such that $e_2(\sigma')$ is still undefined. By monotonicity, $e_1(\sigma') \sqsubseteq e_1(\sigma)$, and hence $e_1(\sigma') \in \kappa$ by Lemma 6. Hence, $e(\sigma') = e_2(\sigma')$ which is undefined. If $e_1(\sigma)$ is defined, but $e_1(\sigma) \notin \kappa$ the reasoning is similar.
- If $e = \{e_2 \mid x \in e_1\}$, then we have three possibilities.
  1. If $e_1(\sigma)$ is undefined, then the result follows from the induction hypothesis.
  2. If $e_1(\sigma)$ is defined, but is not a set, then let $\sigma' = \min(\sigma)$. By monotonicity, $e_1(\sigma')$ cannot be a set, and hence $e(\sigma')$ is undefined. Moreover, $\sigma' \in \mathcal{E}_{c(e_1,1)} \subseteq \mathcal{E}_{c(e,1)}$.
  3. Otherwise, $e_1(\sigma)$ is defined and a set, but there is some $v \in e_1(\sigma)$ such that $e_2(x : v, \sigma)$ is undefined. By the induction hypothesis, there exists $x : u, \sigma_2 \in \mathcal{E}_{c(e_2,1)}$ with $x : u, \sigma_2 \sqsubseteq x : v, \sigma$ such that $e_2(x : u, \sigma_2)$ is undefined. Then $\{u\} \in \mathcal{V}_{\max(1,c(e_2,1))}$, and $\{u\} \sqsubseteq e_1(\sigma)$. By Lemma 9 there exists $\sigma_1 \in \mathcal{E}_{c(e1,\max(1,c(e_2,1)))}$ with $\sigma_1 \sqsubseteq \sigma$ such that $\{u\} \sqsubseteq e_1(\sigma_1)$. Let $\sigma' = \sigma_1 \sqcup \sigma_2$. Note that $\sigma_1 \sqsubseteq \sigma'$ and $\sigma_2 \sqsubseteq \sigma'$. By monotonicity $\{u\} \sqsubseteq e_1(\sigma')$. Hence, there exists some $u' \in e_1(\sigma')$ such that $u \sqsubseteq u$. Then $x : u, \sigma_2 \sqsubseteq x : u', \sigma'$, and $e_2(x : u', \sigma')$ is undefined by monotonicity of undefinedness. Hence, $e(\sigma')$ is undefined. Moreover, $\sigma' \in \mathcal{E}_{c(e_1,\max(1,c(e_2,1))+c(e_2,1))} = \mathcal{E}_{c(e,1)}$. □

**Corollary 7 (Small model for undefinedness).** *Let $e$ be an expression, let $\Gamma$ be a type assignment on $e$, and let $\sigma$ be an environment compatible with $\Gamma$ such that $e(\sigma)$ is undefined. There exists a natural number $l$ which depends only on $e$, and an environment $\sigma' \in \mathcal{E}_l$ compatible with $\Gamma$ such that $e(\sigma')$ is also undefined.*

*Proof.* By Lemma 10, there exists $\sigma' \in \mathcal{E}_{c(e,k)}$ such that $\sigma' \sqsubseteq \sigma$ and $e(\sigma')$ is undefined. Moreover, $\sigma' \in \Gamma$ since $\sigma \in \Gamma$ by Lemma 7. □

Functions mapping atoms to atoms can naturally be lifted to functions defined on values by extending them element- or components-wise. Indeed, let $f$ be a function mapping atoms to atoms, then $f$ can be lifted to values by taking $f((v, w)) = (f(v), f(w))$, and $f(\{v, \dots, w\}) = \{f(v), \dots, f(w)\}$. Also, functions defined on values can be lifted to functions on environments by taking $f(\sigma)(x) = f(\sigma(x))$.

It is easy to show by induction that the PENRC is *generic.*

**Lemma 11 (Genericity).** *Let $e$ be an expression, let $\sigma$ be a environment on $FV(e)$, and let $\rho$ be a permutation of $\mathcal{A}$. If $e(\sigma)$ is defined, then $e(\rho(\sigma)) = \rho(e(\sigma))$. If $e(\sigma)$ is undefined, then so is $e(\rho(\sigma))$.*

**Theorem 3.** *Let $e$ be an expression and let $\Gamma$ be a type assignment on $FV(e)$. It is decidable to check whether $e$ is well-defined under $\Gamma$.*

*Proof.* Suppose that $e$ is not well-defined under $\Gamma$. Then there exists some $\sigma$ compatible with $\Gamma$ such that $e(\sigma)$ is undefined. By Lemma 10 there exists some $\sigma' \in \mathcal{E}_{c(e,1)}$ compatible with $\Gamma$ with $\sigma' \sqsubseteq \sigma$ such that $e(\sigma')$ is undefined.

Let $k$ be a natural number, and let $\tau$ be a type. Let us denote the maximum number of atoms a value in $\tau \cap \mathcal{V}_k$ can mention by $rank(\tau \cap \mathcal{V}_k)$. Then

$$\begin{aligned} rank(\mathcal{A} \cap \mathcal{V}_k) &= 1 \\ rank((\tau_1 \times \tau_2) \cap \mathcal{V}_k) &= rank(\tau_1 \cap \mathcal{V}_k) + rank(\tau_2 \cap \mathcal{V}_k) \\ rank((\tau_1 \cup \tau_2) \cap \mathcal{V}_k) &= \max(rank(\tau_1 \cap \mathcal{V}_k), rank(\tau_2 \cap \mathcal{V}_k)) \\ rank(\mathbf{coll}(\tau') \cap \mathcal{V}_k) &= k\, rank(\tau' \cap \mathcal{V}_k) \end{aligned}$$

Consequently, the maximum number of atoms an environment in $\mathcal{E}_{c(e,1)}$ compatible with $\Gamma$ can mention is bounded by

$$l = \sum_{x \in FV(e)} rank(\Gamma(x) \cap \mathcal{V}_{c(e,1)}).$$

Note that $l$ depends only on $e$ and $\Gamma$. Let $A = \{a_1, \ldots, a_l\} \subseteq \mathcal{A}$. Let us denote the atoms occurring in an environment $\delta$ by $\mathcal{A}(\delta)$. Since the number of different atoms occurring in $\sigma'$ is at most $l$, there exists a permutation $\rho$ of $\mathcal{A}$ such that $\mathcal{A}(\rho(\sigma')) = \rho(\mathcal{A}(\sigma')) \subseteq A$. By genericity, $e(\rho(\sigma'))$ is also undefined.

Hence, in order to check if $e$ is well-defined under $\Gamma$ it suffices to enumerate all environments $\gamma$ compatible with $\Gamma$ that mention only atoms in $A$, and check whether $e(\gamma)$ is defined. There are only a finite number of such $\gamma$, from which the result follows. □

**Lemma 12.** *If $v$ is a value and $v \notin \tau$, then there exists a natural number $k$ and $u \in \mathcal{V}_k$ such that $u \sqsubseteq v$ and $u \notin \tau$.*

*Proof.* Let us define the complexity $c(\tau')$ of a type $\tau'$ as follows.

$$\begin{aligned} c(\mathcal{A}) &= 0 \\ c(\tau_1 \times \tau_2) &= \max(c(\tau_1), c(\tau_2)) \\ c(\tau_1 \cup \tau_2) &= c(\tau_1) + c(\tau_2) \\ c(\mathbf{coll}(\tau')) &= \max(1, c(\tau')) \end{aligned}$$

Let $v$ be a value with $v \notin \tau$. We show that there exists a value $u \in V_{c(\tau)}$ with $u \sqsubseteq v$ such that $u \notin \tau$ by induction on $\tau$.

- If $\tau = \mathcal{A}$, then take $u = v$.
- If $\tau = (\tau_1, \tau_2)$, then $v = (v_1, v_2)$ and either $v_1 \notin \tau_1$, or $v_2 \notin \tau_2$. We can apply the induction hypothesis in both cases.

– If $\tau = \tau_1 \cup \tau_2$, then $v \notin \tau_1$ and $v \notin \tau_2$. By the induction hypothesis there exist $u_1 \in c(\tau_1)$ and $u_2 \in c(\tau_2)$ with $u_1 \sqsubseteq v$ and $u_2 \sqsubseteq v$ such that $u_1 \notin \tau_1$ and $u_2 \notin \tau_2$. Take $u = u_1 \sqcup u_2$, and suppose $u \in \tau$. Then either $u \in \tau_1$, or $u \in \tau_2$. If $u \in \tau_1$, then also $u_1 \sqsubseteq u$ would have to be in $\tau_1$ by Lemma 7, which is a contradiction. If $u \in \tau_2$, then also $u_2 \sqsubseteq u$ would have to be in $\tau_2$, which is also a contradiction. Hence, $u \notin \tau$. Moreover, $u \in V_{c(\tau_1)+c(\tau_2)} = V_{c(\tau)}$.
– Finally, if $\tau = \mathbf{coll}\, \tau')$, then there exists some $v' \in v$ such that $v' \notin \tau'$. By the induction hypothesis there exists $u' \in V_{c(\tau')}$ such that $u' \sqsubseteq v'$ and $u' \notin \tau'$. Then $\{u'\} \sqsubseteq v$ and $\{u'\} \notin \tau$. □

**Corollary 8 (Small model for semantic type-checking).** *Let $e$ be an expression, let $\Gamma$ be a type assignment on $e$ such that $e$ is well-defined under $\Gamma$, and let $\tau$ be a type. Let $\sigma$ be an environment on $e$ such that $e(\sigma) \notin \tau$. There exists a natural number $l$ which depends only on $e$ and $\tau$, and an environment $\sigma' \in \mathcal{E}_l$ compatible with $\Gamma$ such that also $e(\sigma') \notin \tau$.*

*Proof.* Since $e(\sigma) \notin \tau$, there exists a natural number $k$ and a value $u \in \mathcal{V}_k$ with $u \sqsubseteq e(\sigma)$ such that $u \notin \tau$ by Lemma 12. By Lemma 9, there exists $\sigma' \in \mathcal{E}_{c(e,k)}$ such that $\sigma' \sqsubseteq \sigma$ and $u \sqsubseteq e(\sigma')$. Since $u \notin \tau$, $e(\sigma')$ is also not in $\tau$ by Lemma 7. Since $\sigma \in \Gamma$, also $\sigma' \in \Gamma$ by Lemma 7. □